\documentclass{llncs}
\usepackage{llncsdoc}
\usepackage[pdftex]{graphicx}
\usepackage{tablefootnote}
\usepackage[hyphens]{url}
\usepackage[]{cite}
\usepackage{amsmath}
\usepackage{natbib}

\begin{document}

\title{Guidelines for Producing Useful Synthetic Data}
\author{Gillian M. Raab \and Beata Nowok \and Chris Dibben} 
\institute{Administrative Data Research Centre - Scotland, University of Edinburgh, School of GeoSciences, Drummond Street, Edinburgh EH8 9XP, UK \\ \email{\{gillian.raab,beata.nowok,chris.dibben\}@ed.ac.uk}}

\maketitle

\begin{abstract}
\noindent
We report on our experiences of helping staff of the Scottish Longitudinal Study to create synthetic extracts that can be released to users. In particular, we focus on how  the synthesis process can be tailored to produce synthetic extracts that will provide users with similar results to those that would be obtained from the original data. We make recommendations for synthesis methods and illustrate how the staff creating synthetic extracts can evaluate their utility at the time they are being produced. We discuss measures of utility for synthetic data and show that one tabular utility measure is exactly equivalent to a measure calculated from a propensity score.  The methods are illustrated by using the R package $synthpop$ to create synthetic versions of data from the 1901 Census of Scotland.\\

\end{abstract}

\section[intro]{Introduction}
This paper describes our practical experiences of providing synthetic data extracts for users of the Scottish Longitudinal Study. We have investigated how different methods of synthesis influence the usefulness of the data, judged by the extent to which analyses of the observed and synthetic data give similar results.  Clearly the disclosure risk (R) associated with synthetic data needs to be considered as well as their utility (U). Releasing the original data would have the highest utility, but also the highest disclosure risk. Ideally one would evaluate synthetic data by considering an R-U trade-off, \citep{statcon}. The focus of this paper is on U rather than R, although we discuss aspects of R in Section \ref{sec:impl}. We propose methods to increase U by tailoring the syntheses to the needs of the user. Whether these methods might also increase R remains to be investigated further.

In Section \ref{sec:LSs} we describe the UK Longitudinal Studies (LSs) and how synthetic data from them will be made available to users. In Section \ref{sec:util} we describe measures of utility for synthetic data. These include a general utility measure and a specific one based on cross-tabulations, both of which have null distributions that can be used to judge the appropriateness of the synthesis models. In Section \ref{sec:eval} we use the historic census data to illustrate our experiences. In Section \ref{sec:guide} we summarise our provisional recommendations to guide staff starting out on a synthesising task on how to produce useful synthetic data sets. Section \ref{sec:conclude} presents our conclusions.

\section[LSs]{The UK Longitudinal Studies}
\label{sec:LSs}
The England and Wales Longitudinal Study (ONS LS) \citep{hat}, the Scottish Longitudinal Study (SLS) \citep{boyle} and the Northern Ireland Longitudinal Study (NILS) \citep{or} are rich microdata sets linking samples from the national census in each country to administrative data for individuals and their immediate families across several decades. All of the LSs have a similar structure. At their core are pseudo-random samples the UK decennial censuses for the relevant country. Individuals are linked over time across censuses and to administrative data on births, deaths, marriages, records of immigration and emigration from the relevant country and other sources. The data are extremely sensitive. The core census data are controlled by the Census Acts which means that public access is not available for 100 years. Inclusion in the studies is by a number of ``secret" birthdays known only to a very few core staff in each study. No resident of the UK knows whether they are included in one of the LSs. The inferences of interest to LS users are those for a hyper-population model that is presumed to have generated the actual data. Thus it can usually be assumed that such data are generated by simple random sampling (SRS) and the synthetic data are generated and analysed by SRS methods. Following an application to use a study the user specifies the data required and an extract is prepared. Access to the microdata is restricted to trained and approved researchers who can only view and work with the data in safe settings.

\subsection{Implementation of synthetic data in the Scottish Longitudinal Study}
\label{sec:impl}
 In Scotland the proposal to supply synthetic data to the SLS users has been accepted and the implementation process is under way. A user receiving synthetic data must sign an agreement to keep it confidential and to destroy it at the end of their project. Synthetic data have been produced and released for  pilot projects and the work we present here was motivated by feedback from them.

Completely synthetic data such as those to be provided to the SLS users do not by definition include real units and might be expected to present a low, though non-zero, disclosure risk. The fact that the birthdays are secret will reduce this disclosure risk \citep{drechsler_reiter_nonparam}. Evaluations of other synthetic data products \citep{kinn_etal, Abowdetal2006, DrechslerBenderRassler2008, Huetal2014,  McClureReiter2016} have judged their disclosure risks to be low as was the case for an investigation of our methodology \citep{Ell}. Nevertheless, the SLS research board requires some additional statistical disclosure control (SDC) to be undertaken to reduce them, including the addition of noise to any continuous variables where individual values may be disclosive as well as the  removal of records that are unique in the original data and also appear as unique in the synthetic data.   \citet{CS} distinguishes disclosure potential from disclosure harm where the latter may be modified by disclosure management practices from the agency. Thus, although there may be some non-zero disclosure potential in the synthetic data, the restricted access to the synthetic extracts and the SDC should reduce the expected disclosure harm. Harm to the agency may result if an outsider views what they consider to be a potential disclosure risk even if there is no real risk. Thus additional labelling measures are taken to make it clear that the data are false \citep{BN_SJIAOS}. 

Users of the SLS can request synthetic data to allow them to develop code that is then run on the original SLS data either by the user at a visit to the safe setting or on the user's  behalf by the staff at the SLS Development and Support Unit (SLS-DSU)\footnote{\url{http://sls.lscs.ac.uk/}}.  The most time-consuming stage of any practical data analysis project is the first stage of data preparation and cleaning that precedes any formal analyses. Users are expected to carry out such tasks themselves, although the SLS-DSU staff may offer advice. This includes the decisions as to how to handle missing observations. Thus initial synthetic extracts are supplied with missing values synthesised using a missing-at-random approach. Synthetic data can be supplied at any stage of a project, most often when users get their initial extract, but also potentially at a later stage after code to clean and reorganise their data, including any imputation of missing values, has been run on the original SLS data.

\subsection{Methods of producing synthetic data}
\label{sec:method}
Our methods for producing synthetic data for the SLS have been described elsewhere \citep{synthpop_JPC,BN_SJIAOS}. They make use of the package $synthpop$ for \textbf{R}  ( \citet{synthpop_JSS}) that has been developed specifically for the SYLLS project for the LSs\footnote{see \url{http://www.lscs.ac.uk/projects/synthetic-data-estimation-for-uk-longitudinal-studies/}}, but is also freely available to other users \citep{synthpop_pkg}. 
Variables are synthesised one-by-one using sequential regression modelling. This means that conditional distributions, from which synthetic values are drawn, are defined for each variable separately. 

We will be concerned here with evaluating synthesis methods that would typically be used for an SLS user. Thus we will consider cases where only a single synthetic data set is generated from models using estimated parameters from fits to the original data (sometimes termed plug-in sampling). The $synthpop$ package can produce multiple synthetic data sets and can generate data from the predictive distribution of the parameters. However, these features are required only in some circumstances when the user wishes to use the synthetic data in the place of the real data to make population inferences. This will not be the case for SLS users who are using the data for exploratory analyses and will run their final analyses on the real data.

The staff member producing synthetic data can control the synthesis process in various ways, where the three main parameters are
\begin{enumerate}
\item {\textbf{Synthesis method(s)} A different method can be specified for each variable.}
	\item {\textbf{Order of variables}} The parameter $visit.sequence$ for the $syn$ function of $synthpop$  determines the order in which the variables are synthesized.
	\item {\textbf{Choice of predictors} This can be made by defining an appropriate predictor selection matrix (parameter $predictor.matrix$ for $syn$). Its entries are checked to ensure that variables can only be predicted by those that have already been synthesised, i.e. they precede the predicted variable in $visit.sequence$.  }
\end{enumerate}

\noindent{Other parameters deal with special situations, for example to ensure certain deterministic relations in the data are maintained. For any real data set there will be a very large number of possible choices from every possible combination of the three main parameters.
Furthermore, every data extract has specific features that will influence the appropriate choice of parameters. We do not expect to find one prescription as to how to use these parameters for every data extract. However, even at this early stage, our experiences with synthesising test data sets and feedback from SLS-DSU staff have provided some guidelines as to best practice to optimise the utility of the data for the user. 
In the rest of this paper we discuss these and illustrate them with data from the 1901 Census of Scotland made available by the Integrated Census Microdata (I-CeM) project\footnote{https://www.essex.ac.uk/history/research/icem/}. These data are not part of the SLS, but they have many of the same features as SLS census data extracts that can make them challenging to synthesise. In particular there are a large number of records (over 350 thousand), some variables with missing values, some with a very large number of categories, and some deterministic relationships between variables. Details of the I-CeM data set used are presented in the Appendix. The one advantage of using the I-CeM data, over the use of an SLS extract from recent censuses, is that we can present micro-data from the original I-CeM data without any concern about disclosing confidential data.
}

The two main problems with generating synthetic extracts from the SLS are the failure of relationships in the original extract to be matched in the synthetic data, i.e. data utility, and the computing resources required to fit models.
 Our feedback on utility from users supplied with the pilot projects has been expressed in terms of differences between proportions in tables comparing the real and synthetic data. Usually the user will have one or more variables of special interest that may at the end of their project become the outcome variables in their analysis, but at this stage they are very far from the process of fitting formal models. In epidemiological journals the first table usually presents the bivariate relationships between outcome variables and potential predictor variables, and this is often the starting point of our users' analyses.  In Section \ref{sec:util} we discuss measures of data utility that are appropriate for evaluating the utility of tables. 
The SLS-DSU staff have access to reasonably powerful PCs, that are more than adequate for analysing user's extracts. But the $synthpop$ package can require large amounts of memory in some cases and can also require long computing times. We will show that, in some cases, steps to reduce the computing resources required can improve the quality of synthetic data. 
\section{Measures of data utility}
\label{sec:util}
\subsection{Graphical methods}
\begin{figure}[t]
	\centering
	\includegraphics[height=2in]{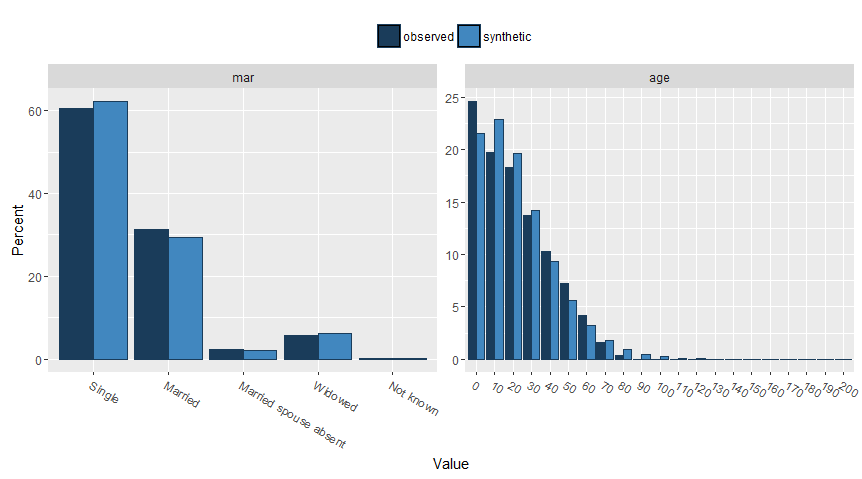}
	\caption[ ]{Sample of univariate comparisons for two variables in the I-CeM data, output from $compare.synds$.
		 Marital status is synthesised from a multinomial logistic regression and age from a square root Normal distribution.} 
	\label{fig1}
\end{figure}

\begin{figure}[h]	
	\centering
	\includegraphics[height=2in]{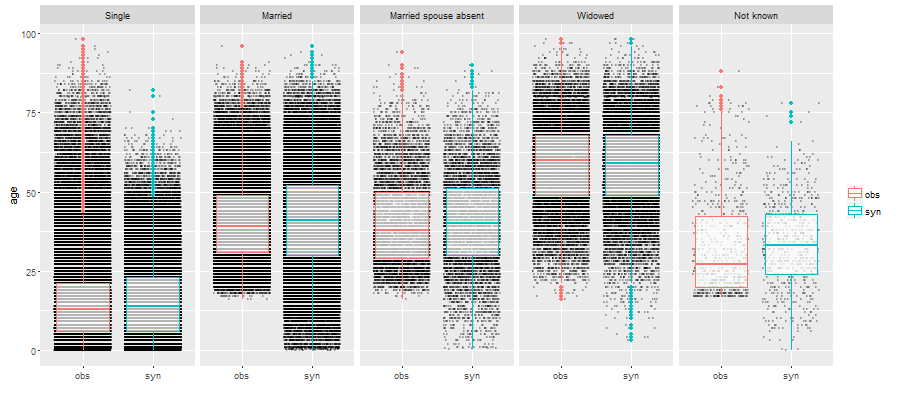}
	\caption[ ]{Bivariate comparison of age by marital status for I-CeM data, output from $multi.compare$. Marital status is synthesised from a multinomial logistic regression and age from a square root Normal distribution.} 
	\label{fig1a}
\end{figure}

A minimum requirement for synthetic data to be useful is that all the marginal distributions of the original and synthetic data should match. This is most easily checked graphically and an example comparing two variables from the I-CeM data is shown in Fig. \ref{fig1}, the context of which will be discussed in Section \ref{sec:meth}. It is clear that there are important differences between the real and synthetic data here, especially for age where a few synthetic people have ages over 100 up to almost 200, and also for marital status where there are too many single people in the synthetic data. Staff synthesising data can and should run these visual comparisons using the $compare.synds$ function in $synthpop$ as a first check on the synthetic data. A more important aspect of data utility is whether the relationships between variables are the same in the original and synthetic data. This can be evaluated either globally, by comparing the two distributions, or by looking at specific interrelationships between certain variables of interest. Visual checks of bivariate comparisons are also available with the $synthpop$ function $multi.compare$ and, if resources allow, this should also be used for all pairs of variables that may be used in inferences. An example is shown in Fig. \ref{fig1a}.

\subsection{Global measures of utility}
The global measure of data utility implemented in the $synthpop$ is derived from a method proposed by  \citet{Wooetal}, which has been standardised specifically for the case of synthetic data (\citep{Joshua}). The real and synthetic data are combined and a propensity score is calculated to predict the source of the record (original or synthetic). A summary measure is then calculated 
as the mean squared difference of the scores from the null value, which is 0.5 when the sample sizes are equal. \citet{Joshua} have shown that  the null distribution of the propensity mean square error ($pMSE$) for completely synthesised data when the model used to generate the synthetic data is correct is $\chi^2_{k-1}/(8N)$, where the real and synthetic data each have $N/2$ records and  $k$ is the number of parameters in the model used to discriminate between the original and synthetic data. We will refer to $8N(pMSE)$ as $U_{gen}$, a general utility measure that will have a $\chi^2_{k-1}$ distribution when the synthesis model is exactly correct.  While this summary statistic may be useful for comparing different approaches to synthesis, it is not, by itself, helpful in diagnosing problems. But the coefficients of the propensity score model can be used to carry out such diagnoses as proposed in \citet{DR2009} and \citet{Dbook}. 
\subsection{Utility from fitted models}
When a statistical model such as a linear or logistic regression has been fitted to synthetic data, the results can be compared with those from the original data. Measures such as the overlap of confidence intervals \citep{Reiter2005, DR2009, K_LBD, NowokHel} or standardised coefficient differences  \citep{RaabNowok2017}.
However, staff producing synthetic data for researchers cannot readily use these measures   because at this stage the researchers  will not usually have formalised the models they intend to fit. It is much ore likely that they can specify the tabulations of interest.
\subsection{Tabular measures}
A $chi^2$ test to compare 
 table entries for the original and synthetic data is an obvious choice of utility measure and has been investigated for synthetic data by \citet{VoasW}. We denote the counts in the $i^{th}$ cell of a tables as  $y_{i}$ from the original data and $s_{i}$ for the synthetic where $i$ ranges over all of the $k$ entries in the tables being compared.
The usual $\chi^2$ statistic to compare the synthesised counts $S$ with the observed, would be $\sum_{i}{(s_i-y_i)^2/y_i}$, which might be expected to have a  $\chi^2$ distribution with degrees of freedom one less than the total cells in the table.  But our situation differs from the usual one where, the $\chi^2$ distribution applies when $y_i$ is the expectation of  $s_i$. We are interested in the case where both $s_i$ and $y_i$ are generated from the same, unknown, distribution. In this case it is possible for $y_i$ to be zero when $s_i$ is not if observations are generated for a combination of variables that do not exist in the observed data. Voas and Williamson have evaluated several options for this situation and recommend a modification due to Neyman (see \citep{RC}) that replaces the $y_i$ in the denominator with $(y_i+s_i)/2$ and where cells with both $s_i$ and $y_i$ zero make no contribution. We denoted this lack-of-fit measure as $U_{tab}$.  Simulations carried out with synthetic data tables including cells with small expected values show that its distribution is well approximated by a  $\chi^2$ distribution with degrees of freedom one less than the number of cells with $(y_i+s_i)>0$. 

We show here that the Voas and Williamson  $\chi^2$ statistic is exactly equivalent to the propensity score measure calculated from a logisitic model including all the interactions included in the table. The propensity score is calculated by combining the rows of the observed and synthetic data and assigning an indicator variable with the value $1$ for synthetic rows and zero for original rows. The propensity score is then calculated from the predicted value of the indicator variable from the other columns. For a table with $k$ cells we denote the counts for the original and synthetic data in the $i_{th}$ cell as $y_i$ and $s_i$.  The data to calculate the propensity score will consist of $k$ groups where the $i^{th}$ group 
will consist of $y_i$ rows from the original data with indicator 0 and  $s_i$ rows from the synthetic data with indicator 1. The predicted value of the indicator variable for all such rows will be $s_i/(s_i + y_i)$ and the propensity score mean square error ($pMSE$) becomes 
\begin{equation*}
pMSE = \frac{1}{N}\sum_{i=1}^k{(s_i + y_i){(s_i/(s_i + y_i) - 0.5)^2}}
\end{equation*}
reducing to
\begin{equation*}
pMSE = \frac{1}{8N}\sum_{i=1}^k{\frac{(s_i - y_i)^2,}{(s_i + y_i)/2}}.
\end{equation*}
The summation here  is just the Voas Williamson statistic showing that it is exactly equivalent to using the null distribution of the $pMSE$ when the synthesis model is correct as proposed by   \citet{Joshua}.

We have found this measure the most practical one for staff to use when creating synthetic data, as we describe below. It is implemented in the \texttt{utility.tab} function in $synthpop$. If the value of $U_{tab}$ indicates a poor synthesis then the discrepancies for individual cells can be examined to diagnose the problem and suggest improvements in the synthesis process. 

\section{Evaluation of synthesis methods with the I-CeM data}
\label{sec:eval}
\subsection{Data description}
Details of the I-CeM data are provided in the Appendix. They consist of 14 variables of which 12 are categorical for over 350 thousand people resident in Midlothian, Scotland in 1901. Several of the categorical variables have a very large number of categories which can cause computational problems. Thus for our initial evaluation of methods for synthesising these data we omit the three categorical variables with the largest number of categories (country of birth and levels 2 and 3 of the occupation codes). We will return to methods for dealing with such data in Section \ref{I:comp}. We will refer to the data set consisting of the remaining 11 variables as I-CeM(11). 
The $synthpop$ package creates indicator variables when missing values of continuous variables are encountered. This applied to the variable for persons per room here. The indicator was synthesised first and the next step synthesised the non-missing values.  The missing values for  \textit{hh\_inactive} were treated as an additional category. 
\subsection{Methods for each conditional distribution}\label{sec:meth}
In common with others we have found methods based on classification and regression trees (CART) \citep{Breiman_CART} to have advantages over parametric models for creating synthetic data. To illustrate this we have synthesised I-CeM(11) in two different ways. The first method uses parametric methods for all variables. The distributions of the two numerical variables were investigated and the transformations that gave fairly symmetric distributions with a reasonable approximation to 
Normality were a square root for age and a cube root for persons per room.
Thus these two variables were synthesised assuming a square-root Normal distribution and a cube root Normal distribution. The categorical variables were synthesised by logistic regression (\textit{sex}) or multinomial regression for the other variables with more than one category. 
 The second method used an implementation of CART for all variables. 
 
 Two implementations of CART are available in \textbf{R}. 
 The first method $rpart$ evaluates every possible split in forming trees and requires a complexity parameter to be set to decide the size of the final tree. 
  For large samples, such as this, the complexity parameter needs to be set to a very small value. If this method is used as a classifier, without cross-validation,  it can result in over-fitting and false-positive results when for syntheses this does not pose a problem because any over-fitting just contributes to the noise in the synthesised data. It has the disadvantage that it tends to favour the inclusion of variables with many categories in the model.
  The alternative method $ctree$ carries out preliminary tests to select important variables, so as to reduce the number of splits that need to be evaluated. We have found both methods to be useful and having a choice is helpful in case a user experiences computational problems.
   For both the parametric and the CART syntheses the variables were ordered as in Table \ref{tab:icem}, except for the occupation code (level 1) and parish which were moved to the end of the synthesis. These were the remaining variables with the largest number of categories so that moving them to the end can help to reduce computational problems, as has also been identified by  \citet{drechsler_reiter_nonparam}. Computing times on a laptop PC were under 10 minutes for CART and between 2 and 3 hours for the parametric synthesis.

      \begin{figure}[t]
      	\centering
      	\includegraphics[height=2in]{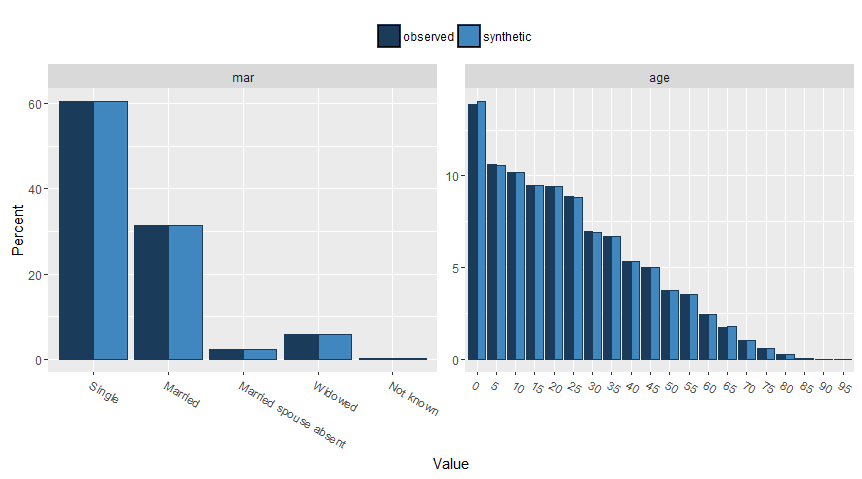}
      	\caption[ ]{Sample of univariate comparisons for two variables in the I-CeM data. Marital status is synthesised from a multinomial logistic regression and age from age transformed to Normal scores.} 
      	\label{fig2}
      \end{figure}

The comparison in Fig. \ref{fig1} shows that the square root Normal fit used in the parametric synthesis was not a good enough fit to the age distribution to make useful synthetic data. This distribution of marital status has also been distorted in the synthetic data by the misfit to age. An alternative parametric method is to predict from Normal scores, and then back-transformed to the percentiles of the original distribution using a method similar to that described by \citet{Woodcock}. When this method was used for the two continuous variables the results were satisfactory, as is shown for age in Fig. \ref{fig2}, where we can see that the fit for marital status is now also satisfactory, as were the fits for all the other categorical variables. The CART synthesis gave satisfactory comparisons for all variables.
    
The $pMSE$ measure, $U_{gen}$, of general utility was calculated for three models. The first model took a bootstrap sample of each variable, without modelling any of the dependencies between variables. The remainder were the  parametric and CART models described above with the numerical variables synthesised from Normal scores. Ideally we would have liked to evaluate a propensity score model with the main effects and interactions of all the variables, but this would have generated a model with 2,789 parameters, beyond the computing power available. Instead we evaluated utility from a model with the main effects of each variable with 85 parameters. Table \ref{tab:PMSE} summarises the results. As expected the bootstrap samples give a very poor result, with $U_{gen}$ over 600 times its expectation. The CART method gives a better $U_{gen}$ than does the parametric synthesis.  
    
    \begin{table}
    	\caption{$pMSE$ based utility measures for different syntheses.}
    	\label{tab:PMSE}
    	\renewcommand{\arraystretch}{1.4}
    	\setlength\tabcolsep{3pt}
    	\centering
    	\begin{tabular}{lrrrl}
    		\hline
    		&   &  &  & Variables with  propensity 	\\	
    		&   $U_{gen}$& df & Ratio  & score Z statistics over 1.7 \\
    		\hline
    		Bootstrap samples & 52,008.18 & 83 & 626.60 & 39 from 84 coefficients\\ 
    		Parametric (1)* & 1,062.55 & 83 & 12.80 & $ age $, $ pperroom $, $ mar $,$ relat $, $ parish $ \\ 
    		Parametric (2)** & 293.86 & 83 & 3.54 & $ mar $, $ relat $, $ age $, $ pperroom $, $ parish $ \\ 
    		Parametric (2) with rules*** & 202.80 & 83 & 2.44 & $ mar $, $ relat $, $ pperroom $, $ parish $ \\ 
    		CART & 155.60 & 83 & 1.87 & $ parish $ \\ 
    		
    		\hline
    		\multicolumn{5}{l}{* Default parametric methods with square root for $ age $ and cube root for $ pperroom $;}\\
    		\multicolumn{5}{l}{** Default parametric methods with Z score transformation for $ age $ and $ pperroom $;}\\
    		\multicolumn{5}{l}{*** Constraint that all people under 16 must be ``Single".}\\
    				
    	\end{tabular}
    \end{table}

 Examination of the coefficients of the fitted propensity score model showed that some of the categories of  marital status and relationship to head of household, as well as age were the most important coefficients of the propensity score. This suggested calculating $U_{tab}$ for these variables and results are in Table \ref{tab:tabutil}. The tables for age were formed by dividing into groups based on  the quintiles of the observed data. The CART models showed no evidence of a lack of fit for any of these tables, while the $U_{tab}$ values show an extremely poor fit.  Examination of the relationship between age and marital status, see Fig. \ref{fig1a}, soon revealed the problems. The marital status of all people under 16 in the original data was ``Single". The CART synthesis conformed to this rule, while over 5,000 observations in the parametric synthesis had age under 16 and a marital status other than ``Single".  This suggested that the parametric synthesis might be rerun with a rule set to constrain all of those under 16 to have marital status ``Single". Results for this model are shown in Tables \ref{tab:PMSE} and \ref{tab:tabutil}.
 
Although $U_{gen}$ is improved, the parameters of the propensity score model still indicate serious problems with both marital status and relationship to head of household, but there was no evidence that there was a problem with age. But looking at the table of $U_{tab}$ values we can see that the tables by age for the parametric models, although improved by imposing the rule, still have very  poor utility. There was no evidence of a lack-of-fit of the CART model.
  
\begin{figure}[h]
    	\centering
    	\includegraphics[height=2in]{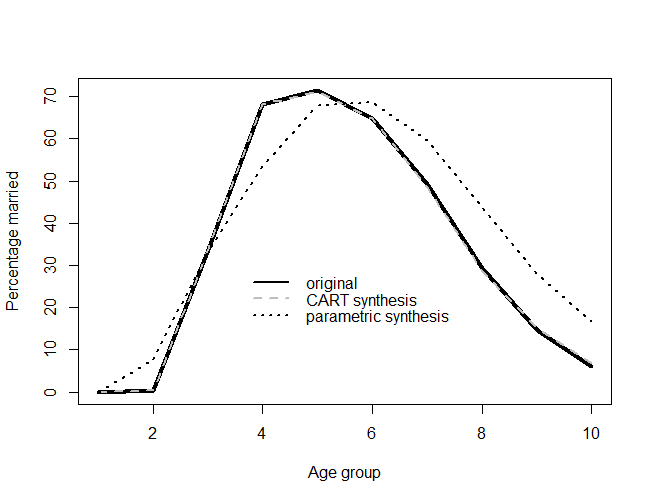}
    	\caption[ ]{Marital status by age group for the original and synthesised I-CeM data. Parametric synthesis constrained those under 16 to marital status ``Single".} 
    	\label{fig3}
    \end{figure}

Examination of the detailed tables showed that the problem arose with the non-linear relationship between marital status and age. Fig. \ref{fig3} shows the percentage married by ten-year age groups for parametric models, with the rule for marital status imposed  and for CART synthesis. The differences between the original data and CART were under 0.5\% in every age group and hence are barely visible in the figure. The relationship for the parametric synthesis was very different. It may seem surprising that the relationship with married status is not linear with age for the parametric model. We investigated whether the non-linear pattern was due to the interaction with sex, which was also in the model that predicted marital status. But changing the ordering of the variables made little difference. The reason for the non-linearity is that ``married" is just one component of a multinomial model. Details of the fitted values are shown in Table \ref{tab:mar} where we can see that there is a lack of fit in all the categories, and the largest category overall (``Single") is enforced to approximate linearity, and the fit to the married category is conditional on this. To overcome this it would be possible to include further terms for age in the parametric synthesis, but in view of the superior performance of the CART models this does not seem worth pursuing here.    
    
     \begin{table}
     	\begin{center}
     		\caption{Tabular utility measures for different tables and synthesis methods.}
     		\label{tab:tabutil}
     		\renewcommand{\arraystretch}{1.4}
     		\setlength\tabcolsep{3pt}
     		\begin{tabular}{llrrr}
     			\hline
     			Method  &  Table  	&    $U_{tab}$   &  df &  Ratio 	\\	
     			\hline 
     			Parametric (1) & $ mar $ by $ relat $   & 1,716.38 & 36 & 47.68 \\ 
     			& $ mar $ by $ age $   & 40,139.64 & 24 & 1,672.49 \\
     			& $ relat $ by $ age $ & 41,698.15 & 39 & 1,069.18 \\  
     			Parametric (2) &  $ mar $ by $ relat $ & 614.38 & 36 & 17.07 \\ 
     			&  $ mar $ by $ age $ & 29,560.16 & 24 & 1,231.67 \\ 
     			&  $ relat $ by $ age $ & 31,237.17 & 39 & 800.95 \\ 
     			Parametric (2) with rules &  $ mar $ by $ relat $ & 259.18 & 36 & 7.20 \\ 
     			&  $ mar $ by $ age $ & 16,491.31 & 20 & 820.97 \\ 
     			&  $ relat $ by $ age $ & 22,749.37 & 38 & 598.67 \\ 
     			CART &  $ mar $ by $ relat$   & 45.42 & 37 & 1.23 \\ 
     			&  $ mar $ by $ age $ & 18.71 & 20 & 0.94 \\ 
     			& $ relat $ by $ age $ & 51.95 & 38 & 1.37 \\ 
     			\hline
     			\multicolumn{5}{l}{\textit{Notes}: For details of methods see Table 1.}\\
     				\end{tabular}
     	\end{center}
     \end{table} 
     
     \begin{table}
     	\caption{Marital status by age group for the original data and data generated by parametric synthesis.}
     	\label{tab:mar}
     	\renewcommand{\arraystretch}{1.4}
     	\setlength\tabcolsep{3pt} 
     	\centering
     \begin{tabular}{rrrrrrrrr}
     	\hline
     	Age 	&\multicolumn{4}{c}{Original data} &  \multicolumn{4}{c}{Parametric synthesis}\\
     	group 	& Single & Married & Other & N & Single & Married & Other & N \\ 
     	\hline
     	0--9   & 100.00 & 0.00 & 0.00 & 79,831 & 100.00 & 0.00 & 0.00 & 79,430 \\ 
     	19--20 & 99.24 & 0.45 & 0.31 & 70,976 & 91.38 & 7.83 & 0.79 & 71,083 \\ 
     	29--30 & 62.16 & 33.85 & 3.98 & 65,252 & 62.94 & 33.37 & 3.70 & 65,042 \\ 
     	39--40 & 24.17 & 68.17 & 7.66 & 50,296 & 39.41 & 53.38 & 7.21 & 50,573 \\ 
     	49--50 & 15.09 & 71.42 & 13.49 & 37,719 & 18.39 & 67.89 & 13.72 & 37,959 \\ 
     	59--60 & 12.03 & 64.81 & 23.16 & 27,002 & 7.60 & 68.60 & 23.80 & 27,014 \\ 
     	69--70 & 11.76 & 49.09 & 39.15 & 16,511 & 2.49 & 59.70 & 37.82 & 16,366 \\ 
     	79--80 & 13.74 & 29.40 & 56.85 & 6,737 & 0.82 & 43.68 & 55.50 & 6,838 \\ 
     	89--90 & 13.77 & 14.60 & 71.63 & 1,438 & 0.21 & 27.81 & 71.98 & 1,449 \\ 
     	99--98 & 15.85 & 6.10 & 78.05 & 82 & 0.00 & 16.67 & 83.33 & 90 \\ 
     	\hline
     \end{tabular}
    \end{table}   
     		        		
\subsection{Ordering of variables}
  
In view of the results in Section \ref{sec:meth} we limit our examples on the ordering of variables to those for CART models. Experience with other data sets suggests that conclusions are roughly similar, although often difficult to distinguish from other aspects of lack-of-fit for parametric models. We noted in Section \ref{sec:meth}
that, although the CART method was better than the parametric method, there was still a substantial lack-of-fit as assessed by $U_{gen}$. The coefficients of the propensity score pointed to the occupation code as being a main contributor to this. We have thus investigated $U_{tab}$ for tables of occupation code with all the other variables. Results are in Table \ref{tab:occord}. Note that the occupation code was the penultimate variable in the visit sequence and parish, with even more categories, was the last one. The CART synthesis was rerun with the occupation code moved to the start of the visit sequence.  The effect of the two orderings on $U_{tab}$ is presented in Table \ref{tab:occord} and shows that the effect of ordering of the variables is fairly modest. When the occupation code is at the start there is some improvement in the utility for the first variables synthesised, but the differences are not great. Experimenting with other orderings showed that, when \textit{hh\_occ1} was the first variable in the sequence, tables with the second and third variables always had good utility, but things could get worse after that, especially for variables, like number of servants and employment status here, with a strong relationships to the first variable.

\begin{table}
\caption{$U_{tab}$ for occupation code (level 1) against other variables.}
\label{tab:occord}
\renewcommand{\arraystretch}{1.4}
\setlength\tabcolsep{3pt} 
\centering
\begin{tabular}{lrrrrrr}
\hline
Variable & \multicolumn{3}{c}{$ hh\_occ1 $ at end} & \multicolumn{3}{c}{$ hh\_occ1 $ at start} \\
 & $U_{tab}$ & df & Ratio & $U_{tab}$ & df & Ratio \\ 
\hline
$ sex $ & 169.9 & 48 & 3.54 & 54.2 & 48 & 1.13 \\ 
$ age $ & 827.4 & 120 & 6.89 & 478.8 & 120 & 3.99 \\ 
$ mar $ & 462.1 & 120 & 3.85 & 334.2 & 120 & 2.78 \\ 
$ relat $ & 1,411.9 & 192 & 7.35 & 644.9 & 192 & 3.36 \\ 
$ disability $ & 477.1 & 156 & 3.06 & 302.1 & 154 & 1.96 \\ 
$ servants $ & 1,456.6 & 95 & 15.33 & 1,039.3 & 93 & 11.18 \\ 
$ hh\_employ $ & 62.6 & 72 & 0.87 & 5,515.2 & 72 & 76.60 \\ 
$ hh\_inactive $ & 2,488.9 & 145 & 17.16 & 1,944.6 & 138 & 14.09 \\ 
$ pperroom $ & 748.3 & 120 & 6.24 & 952.3 & 120 & 7.94 \\ 
$ parish $ & 17,711.0 & 661 & 26.79 & 86,143.0 & 691 & 124.66 \\ 
\hline
\end{tabular}
\end{table}

\subsection{Selecting predictor variables}
In the previous section we have fitted models for each variable and generated synthetic data using all the variables preceding the variable being synthesised.
We have not found it necessary to define reduced models for CART, since the procedure already selects predictors. Such an approach may be more useful for parametric models, where computational problems due to sparse data may arise. A better solution to dropping predictors here might be to use some method such as ridge regression to fit multinomial models, though we have not found it necessary for any of the syntheses we have investigated for the I-CeM data or other data sets we have investigated. An exception to fitting models with all the available variables arises from the need to synthesise variables that have many categories. This is covered in the next section. 
\subsection{Computational issues}\label{I:comp}
Three variables were omitted from the analyses presented above. These were country of birth (77 levels) and the two finer occupation codes, level 2 (76 levels) and level 3 (711 levels), where these are nested within their grouped categories. Variables like this are often requested by SLS users, but it is not feasible to model them with the computing resources available.  It is unlikely that a user would wish to present any data based on such variables in a final analysis, but they might wish to investigate them so as to form new groupings by recoding them. In order to provide some data to users for this purpose we suggest the following approach.
\begin{enumerate}
	\item{Form a grouping into larger categories if one is not available.}
	\item{Use the grouped variable in the synthesis of all the other variables.}
	\item{After all the other variables have been synthesised, synthesize the variable with a large number of categories from ONLY its grouped variable by selecting a bootstrap sample within each grouped category with the $syn.nested$ function in $synthpop$.}
\end{enumerate}

This approach was adopted for the three variables with large numbers of categories in the I-CeM data. The occupation variables required no recoding as they were already nested. Country of birth was coded to the four largest categories and the other smaller groups pooled together. This approach produces useful data, but it needs to be explained to the user that the variables with many categories should not be used except for univariate analyses.

A further computational problem is that of very large data sets that can cause \textbf{R} to run out of memory. This can be obviated by stratifying the data into smaller subgroups and synthesising within subgroups. As well as reducing the computational burden this approach guarantees that tables of the stratifying variable and any other variable, or combination of variables, will have good $U_{tab}$. When the I-CeM data were stratified by \textit{hh\_occ1} the $U_{tab}$ values, as shown in Table \ref{tab:occord} all had ratios around 1 with no evidence of a lack-of-fit. Thus this move, undertaken for computational reasons, also had benefits for the utility of the relationships between the stratifying variable and other variables.
\section{Provisional guidelines and recommendation}
\label{sec:guide}
The following recommendations are drafted for the benefit of SLS-DSU staff who are creating synthetic data for users. 
\begin{enumerate}
	\item{Use a method based on CART for all of the variables to be synthesised.}
	\item{Define any constraints you are aware of by setting up rules. Although this may not be needed for CART models it should always be done since we cannot be sure that the CART models will always respect the constraints. It will not hurt and may reduce computing time.}
	\item{Check for any variables with very large numbers of categories and consider grouping them as described in Section \ref{I:comp}.}
	\item{Discuss with the user what they might consider to be their major outcome variable or variables, and if numbers in subgroups are large enough consider stratifying on them.}
	\item{Keep variables whose inter-relationships are important together near the start of the synthesis, though this may be of minor importance.}
	\item{Move variables with many categories to the end of the synthesis, although this may conflict with the previous point.}
	\item{Once the synthesis is complete inspect the univariate comparisons between the original and synthesised data along with bivariate comparisons considered of particular interest.}
	\item{Check on measures of general utility, $U_{gen}$, perhaps taking subgroups of variables to help interpretation.}
	\item{Follow up any problematic $U_{gen}$ results by calculating $U_{tab}$ for tables suggested by the coefficients of the propensity score.}
	\item{Remember that these are only provisional guidelines and be ready to try different or new methods.}
	
\end{enumerate}
\section{Conclusions}
\label{sec:conclude}
We have set out provisional recommendations on how to create synthetic data and, in particular, how to evaluate their utility and diagnose difficulties. The constraints of computing power have been a limiting factor in how we have been able to synthesise data. In future we expect that these may be lessened by the availability of better equipment. We have written our code in the very flexible \textbf{R} language, which is excellent for developing methods, but not computationally efficient. If our methods become more set in stone, it may be worthwhile to code some of the routines in another language.

Our method of creating synthetic data is different from, and more challenging than, the way in which other synthetic data products have been created. Usually a single data set is made available for all potential users and may be created by a team of people who use methods that are tuned to the specific data set and are evaluated thoroughly over a period of time. In our case we are creating tools to allow synthesisers in the SLS-DSU unit to produce usable data sets in a fairly tight time frame to allow users to explore their extracts. Many of the methods we have found useful are those that have also been used by the teams creating single synthetic data products e.g. \citet{SIPP, kinn_etal}. In particular large data sets can benefit from stratification into subgroups. This is often done by geographic area. We have, so far, had relatively little experience of using stratification in the creation of synthetic data. But one recommendation we have found helpful is to stratify longitudinal data according to response patterns (e.g. grouping subjects by the censuses in which they were enumerated). Further research is needed to find out how best to stratify data without making the task of the staff who create the syntheses too complex. This is just one of many aspects of practical synthesis where further research would be beneficial. These include the development of measures of  disclosure risk that can be used routinely to evaluate synthetic data sets.

\section{Acknowledgements}
We acknowledge the help of the staff of the SLS-DSU staff in implementing the use of the $synthpop$ package to create user extracts. Beata Nowok and Gillian Raab are funded by the UK Economic and Social Research Council’s Administrative Data Research Centre – Scotland, Economic and Social Research Centre  Grant ES/L007487/1.

\bibliographystyle{chicago}
\bibliography{Prac_bespoke_syn}

\begin{thebibliography}{}

\bibitem[\protect\citeauthoryear{Abowd, Hawala, Ricchetti, and Stinson}{Abowd
  et~al.}{2006}]{Abowdetal2006}
Abowd, J., S.~Hawala, B.~Ricchetti, and M.~Stinson (2006).
\newblock Testing disclosure risk in the proposed sipp-irs-ssa public use
  files.
\newblock Document submitted to the Census Bureau's Disclosure Review Board on
  November 16, 2006 Available from:
  \url{https://www2.vrdc.cornell.edu/news/wp-content/uploads/2007/11/drbmemonov2006.pdf}.

\bibitem[\protect\citeauthoryear{Abowd, Stinson, and Benedetto}{Abowd
  et~al.}{2006}]{SIPP}
Abowd, J., M.~Stinson, and G.~Benedetto (2006).
\newblock Final report to the social security administration on the
  sipp/ssa/irs public use file project. technical report, longitudinal
  employer-household dynamics program , us bureau of the census.

\bibitem[\protect\citeauthoryear{Boyle, Feijten, Feng, Hattersley, Huang,
  Nolan, and Raab}{Boyle et~al.}{2009}]{boyle}
Boyle, P., P.~Feijten, Z.~Feng, L.~Hattersley, Z.~Huang, J.~Nolan, and G.~M.
  Raab (2009).
\newblock {Cohort Profile: the Scottish Longitudinal Study (SLS)}.
\newblock {\em International Journal of Epidemiology\/}~{\em 38}, 385--392.

\bibitem[\protect\citeauthoryear{Breiman~L}{Breiman~L}{1984}]{Breiman_CART}
Breiman~L, Friedman~JH, O. R. S.~C. (1984).
\newblock {\em Classification and Regression Trees.}
\newblock Wadsworth,Belmont (CA).

\bibitem[\protect\citeauthoryear{Drechsler}{Drechsler}{2011}]{Dbook}
Drechsler, J. (2011).
\newblock {\em Synthetic Data Sets for Statistical Disclosure Control}.
\newblock New York: Springer Science+Business Media.

\bibitem[\protect\citeauthoryear{Drechsler, Bender, and Rassler}{Drechsler
  et~al.}{2008}]{DrechslerBenderRassler2008}
Drechsler, J., S.~Bender, and S.~Rassler (2008).
\newblock Comparing fully and partially synthetic datasets for statistical
  disclosure control in the german iab establishment panel.
\newblock {\em Transactions on Data Privacy\/}~{\em 1}, 105--130.

\bibitem[\protect\citeauthoryear{Drechsler and Reiter}{Drechsler and
  Reiter}{2009}]{DR2009}
Drechsler, J. and J.~P. Reiter (2009).
\newblock Disclosure risk and data utility for partially synthetic data: An
  empirical study using the german iab establishment survey.
\newblock {\em Journal of Official Statistics\/}~{\em 25}, 589–603.

\bibitem[\protect\citeauthoryear{Drechsler and Reiter}{Drechsler and
  Reiter}{2011}]{drechsler_reiter_nonparam}
Drechsler, J. and J.~P. Reiter (2011).
\newblock An empirical evaluation of easily implemented, nonparametric methods
  for generating synthetic datasets, computational statistics and data
  analysis.
\newblock {\em Computational Statistics and Data Analysis\/}~{\em 55},
  3232--3243.

\bibitem[\protect\citeauthoryear{Duncan, Elliot, and Salazar-González}{Duncan
  et~al.}{2011}]{statcon}
Duncan, G.~T., M.~Elliot, and J.~Salazar-González (2011).
\newblock {\em Statistical Confidentiality Principles and Practice}.
\newblock Springer, New York.

\bibitem[\protect\citeauthoryear{Elliot}{Elliot}{2014}]{Ell}
Elliot, M. (2014).
\newblock {Final report on the disclosure risk associated with the synthetic
  data produced by the SYLLS team}.
\newblock
  \url{http://www.cmist.manchester.ac.uk/research/publications/reports/}.

\bibitem[\protect\citeauthoryear{Hattersley and Cresser}{Hattersley and
  Cresser}{1995}]{hat}
Hattersley, L. and R.~Cresser (1995).
\newblock {\em The Longitudinal Study, 1971--1991: History, organisation and
  quality of data. LS Series no.7}.
\newblock London: The Stationery Office.

\bibitem[\protect\citeauthoryear{Hu, Reiter, and Wang}{Hu
  et~al.}{2014}]{Huetal2014}
Hu, J., J.~Reiter, and Q.~Wang (2014).
\newblock Disclosure risk evaluation for fully synthetic data.
\newblock {\em Privacy in Statistical Databases, Lecture Notes in Computer
  Science\/}, 185--199.

\bibitem[\protect\citeauthoryear{Kinney, Reiter, Reznek, Miranda, J., and
  Abowd}{Kinney et~al.}{2011}]{K_LBD}
Kinney, S.~K., J.~P. Reiter, A.~P. Reznek, Miranda, R.~S. J., Jarmin, and J.~M.
  Abowd (2011).
\newblock Towards unrestricted public use business microdata: The synthetic
  longitudinal business database.
\newblock {\em International Statistical Review\/}~{\em 79}, 363 -- 384.

\bibitem[\protect\citeauthoryear{Kinney, Reiter, Reznek, Miranda, Jarmin, and
  Abowd}{Kinney et~al.}{2011}]{kinn_etal}
Kinney, S.~K., J.~P. Reiter, A.~P. Reznek, J.~Miranda, R.~S. Jarmin, and J.~M.
  Abowd (2011).
\newblock Towards unrestricted public use business microdata: The {Synthetic}
  {Longitudinal} {Business} {Database}.
\newblock {\em International Statistical Review\/}~{\em 79\/}(3), 362--384.

\bibitem[\protect\citeauthoryear{McClure and Reiter}{McClure and
  Reiter}{2016}]{McClureReiter2016}
McClure, D. and J.~Reiter (2016).
\newblock Assessing disclosure risks for synthetic data with arbitrary intruder
  knowledge.
\newblock {\em Statistical Journal of the International Association for
  Official Statistics\/}~{\em 32}, 109--126.

\bibitem[\protect\citeauthoryear{Nowok}{Nowok}{2015}]{NowokHel}
Nowok, B. (2015).
\newblock Utility of synthetic microdata generated using tree-based methods.
\newblock Joint UNECE/Eurostat work session on statistical data confidentiality
  7th October Available from
  \url{http://www1.unece.org/stat/platform/display/SDCWS15/} Accessed:
  2016-04-05.

\bibitem[\protect\citeauthoryear{Nowok, Raab, and Dibben}{Nowok
  et~al.}{2015}]{synthpop_pkg}
Nowok, B., G.~M. Raab, and C.~Dibben (2015).
\newblock {\em synthpop: Generating synthetic versions of sensitive microdata
  for statistical disclosure control}.
\newblock R package version 1.1-0.

\bibitem[\protect\citeauthoryear{Nowok, Raab, and Dibben}{Nowok
  et~al.}{2016}]{synthpop_JSS}
Nowok, B., G.~M. Raab, and C.~Dibben (2016).
\newblock {synthpop : Bespoke creation of synthetic data in R}.
\newblock {\em Journal of statistical software\/}~{\em 74}, 1--26.
\newblock Available from \url{https://www.jstatsoft.org/article/view/v074i11}
  Accessed: 2016-11-22.

\bibitem[\protect\citeauthoryear{Nowok, Raab, and Dibben}{Nowok
  et~al.}{2017}]{BN_SJIAOS}
Nowok, B., G.~M. Raab, and C.~Dibben (2017).
\newblock Providing bespoke synthetic data for the {UK} {Longitudinal}
  {Studies} and other sensitive data with the synthpop package for {R}.
\newblock {\em Statistical Journal of the IAOS\/}~{\em Preprint, 2017}, 1--12.

\bibitem[\protect\citeauthoryear{O'Reilly, Rosato, Catney, Johnston, and
  Brolly}{O'Reilly et~al.}{2011}]{or}
O'Reilly, D., M.~Rosato, G.~Catney, F.~Johnston, and M.~Brolly (2011).
\newblock {Cohort description: The Northern Ireland Longitudinal Study (NILS)}.
\newblock {\em International Journal of Epidemiology\/}~{\em 41}, 634--641.

\bibitem[\protect\citeauthoryear{Raab and Nowok}{Raab and
  Nowok}{2017}]{RaabNowok2017}
Raab, G. and B.~Nowok (2017).
\newblock Inference from synthetic data.
\newblock Package vignette for the synthpop package Available at
  \url{https://cran.r-project.org/web/packages/synthpop/vignettes/inference.pdf}.
\newblock Accessed: November 2017.

\bibitem[\protect\citeauthoryear{Raab, Nowok, and Dibben}{Raab
  et~al.}{2017}]{synthpop_JPC}
Raab, G.~M., B.~Nowok, and C.~Dibben (2017).
\newblock Practical data synthesis for large samples.
\newblock {\em Journal of Privacy and Confidentiality\/}~{\em 7\/}(3), 67--97.

\bibitem[\protect\citeauthoryear{Read and Cressie}{Read and Cressie}{1988}]{RC}
Read, T. and N.~Cressie (1988).
\newblock {\em Goodness-of-Fit Statistics for Discrete Multivariate Data}.
\newblock Springer, New York.

\bibitem[\protect\citeauthoryear{Reiter}{Reiter}{2005}]{Reiter2005}
Reiter, J.~P. (2005).
\newblock Releasing multiply imputed, synthetic public use microdata: An
  illustration and empirical study.
\newblock {\em Journal of the Royal Statistical Society Series A: Statistics in
  Society\/}~{\em 168}, 185–205.

\bibitem[\protect\citeauthoryear{Skinner}{Skinner}{2011}]{CS}
Skinner, C. (2011).
\newblock Statistical disclosure risk: seperating potential and harm.
\newblock {\em International Statistical Review\/}~{\em 80\/}(3), 349--368.

\bibitem[\protect\citeauthoryear{Snoke, Raab, Nowok, Dibben, and
  Slavkovic}{Snoke et~al.}{2017}]{Joshua}
Snoke, J., G.~M. Raab, B.~Nowok, C.~Dibben, and A.~Slavkovic (2017).
\newblock General and specific utility measures for synthetic data.
\newblock {\em Submitted\/}.

\bibitem[\protect\citeauthoryear{Voas and Williamson}{Voas and
  Williamson}{2001}]{VoasW}
Voas, D. and P.~Williamson (2001).
\newblock Evaluating goodness-of-fit measures for synthetic microdata.
\newblock {\em Geographical and Environmental Modelling\/}~{\em 5}, 177--200.

\bibitem[\protect\citeauthoryear{Woo, Reiter, Oganian, and Karr}{Woo
  et~al.}{2009}]{Wooetal}
Woo, M.-J., J.~P. Reiter, A.~Oganian, and A.~F. Karr (2009).
\newblock Global measures of data utility for microdata masked for disclosure
  limitation.
\newblock {\em Journal of Privacy and Confidentiality\/}~{\em 1}, 111--124.

\bibitem[\protect\citeauthoryear{Woodcock and Benedetto}{Woodcock and
  Benedetto}{2009}]{Woodcock}
Woodcock, S.~D. and G.~Benedetto (2009).
\newblock Distribution-preserving statistical disclosure limitation.
\newblock {\em Computational Statistics and Data Analysis\/}~{\em 53},
  4228--–4242.

\end{thebibliography}

\section{Appendix}

\begin{table}
\caption{Variables in the I-CeM data used in this paper}
\label{tab:icem}
\renewcommand{\arraystretch}{1.4}
\setlength\tabcolsep{3pt}
\centering
\scriptsize 
\begin{tabular}{lllrrrr}
	\hline
	 No & Variable & Variable & \multicolumn{2}{c}{Missing} & Distinct & Description\\ 
	    & name     & type     & Number & Percentage         & values   &  \\ 
	\hline
	1 & $ parish $ &   categorical &   0 &  &  29 &  Civil parish of registration \\
	2 & $ sex $ &   categorical &   0 &  &   2 & Male or female \\ 
	3 & $ age $ &   numeric &   0 &  & 129 & Age in years (part years if \textless 1) \\ 
	4 & $ mar $ &   categorical &   0 &  &   5 & Marital status \\ 
	5 & $ relat $ &   categorical &   0 &  &   8 &  Relationship to head of household\\ 
	6 & $ disability $ &   categorical &   0 &  &   7 &  Categories of disability\\ 
	7 & $ ctrybirth $ &   categorical &   0 &  &  77 & Country of birth \\ 
	8 & $ nservants $ &   categorical &   0 &  &   4 &  Number of servants in household\\ 
	9 & $ hh\_occ1 $ &   categorical &   0 &  &  24 &  Occupation code (level 1)\\ 
	10 & $ hh\_occ2 $ &  categorical &   0 &  &  76 &  Occupation code (level 2)\\ 
	11 & $ hh\_occ3 $ &  categorical &   0 &  & 711 & Occupation code (level 3)\\ 
	12 & $ hh\_employ $ &  categorical &   0 &  &   3 & Employer or Worker or blank\\ 
	13 & $ hh\_inactive $ &  categorical & 25,594 & 7.2 &   8 & Reason for being inactive \\ 
	14 & $ pperroom $ &  numeric & 3,867 & 1.1 & 271 & Persons per room in household \\ 
	\hline
\end{tabular}
\end{table}

The I-CeM data is an extract from the 1901 Census of Scotland consisting of all persons enumerated in private households in the historic county of Midlothian, which includes the City of Edinburgh, total sample size 355,844.  The variable \textit{disability} is very sparse with only 93 people reported as being deaf, 95 with physical disabilities and the largest disabled category ``Idiot or insane" reported by only 285 people. The variables with names starting with \textit{hh} are derived variables from the employment of the head of household and \textit{pperroom} and \textit{nservants} (coded as none, one, two or more than two) are also household measures. The three occupation codes are hierarchical with level 2 nested within level 1 and level 3 nested within level 2. The level 3 code has many very small categories. For example, there were only single instances of eleven occupations, including just one pin manufacturer and one sword and bayonet maker.

\end{document}